\documentclass[10pt]{article}
\usepackage[T1]{fontenc}
\usepackage{mathptmx}

\usepackage{graphicx}
\usepackage{dcolumn}
\usepackage{bm}
\usepackage{breqn}

\usepackage{indentfirst,csquotes}

\usepackage{amssymb,amsthm,amsmath}
\usepackage{xcolor,paralist,hyperref,titlesec,fancyhdr,etoolbox}

 \hypersetup{ colorlinks=true, linkcolor=black, filecolor=black, urlcolor=black, citecolor=black }

\usepackage{authblk}

\begin{document}

\title{Modeling of Light Production in Inorganic Scintillators} 

\author[1]{B. Kreider}
\author[1]{I. Cox}%
\author[1]{R. Grzywacz}%
\author[2]{J. M. Allmond}
\author[3]{A. Augustyn}
\author[1]{N. Braukman}
\author[4]{P. Brionnet}
\author[5]{A. Esmaylzadeh}
\author[5]{J. Fischer} 
\author[4]{N. Fukuda}
\author[6]{G. Garcia De Lorenzo}
\author[4]{S. Go}
\author[7]{S. Hanai}
\author[1]{D. Hoskins}
\author[7]{N. Imai}
\author[2]{T. T. King}
\author[7]{N. Kitamura}
\author[8]{K. Kolos}
\author[9]{A. Korgul}
\author[9]{C. Mazzocchi}
\author[4]{S. Nishimura}
\author[10]{K. Nishio}
\author[4]{V. Phong}
\author[2]{T. Ruland}
\author[2]{K. P. Rykaczewski}
\author[9]{A. Skruch}
\author[1]{Z. Y. Xu}
\author[7]{R. Yokoyama}

 \affil[1]{\textit{Department of Physics and Astronomy,\ University\ of\ Tennessee,\ Knoxville,\ Tennessee\ 37996,\ USA}}
 \affil[2]{\textit{Physics\ Division,\ Oak\ Ridge\ National\ Laboratory,\ Oak\ Ridge,\ Tennessee\ 37831,\ USA}}
 \affil[3]{\textit{National\ Centre\ for\ Nuclear\ Research,\ Warsaw\ PL\textit{-}02\textit{-}093,\ Poland}}
 \affil[4]{\textit{RIKEN\ Nishina\ Center,\ 2\textit{-}1\ Hirosawa,\ Wako,\ Saitama\ 351\textit{-}0106,\ Japan}}
 \affil[5]{\textit{Institut\ f\textit{\"u}r\ Kernphysik,\ Universit\textit{\"a}t\ zu\ K\textit{\"o}ln,\ 50937\ K\textit{\"o}ln,\ Germany}}
 \affil[6]{\textit{Facultad\ de\ Ciencias\ Físicas,\ Universidad\ Complutense\ de\ Madrid,\ E\textit{-}28040\ Madrid,\ Spain}}
  \affil[7]{\textit{Center\ for\ Nuclear\ Study,\ University\ of\ Tokyo,\ 2\textit{-}1\ Hirosawa,\ Wako,\ Saitama\ 351\textit{-}0198,\ Japan}}
 \affil[8]{\textit{Lawrence\ Livermore\ National\ Laboratory,\ Livermore,\ California\ 94550,\ USA}}
 \affil[9]{\textit{Faculty\ of\ Physics,\ University\ of\ Warsaw,\ Warsaw\ PL\textit{-}02\textit{-}093,\ Poland}}
 \affil[10]{\textit{Advanced\ Science\ Research\ Center,\ Japan\ Atomic\ Energy\ Agency,\ Tokai,\ Ibaraki\ 319\textit{-}1195,\ Japan}}

\date{\today}
\maketitle


\begin{abstract}
In recent experiments, inorganic scintillators have been used to study the decays of exotic nuclei, providing an alternative to silicon detectors and enabling measurements that were previously impossible. However, proper use of these materials requires us to understand and quantify the scintillation process. In this work, we propose a framework based on that of Birks [Proc. Phys. Soc. A 64, 874] and Meyer and Murray [Phys. Rev. 128, 98] to model the light output of inorganic scintillators in response to beams of energetic heavy ions over a broad range of energies. Our model suggests that, for sufficiently heavy ions at high energies, the majority of the light output is associated with the creation of delta electrons, which are induced by the passage of the beam through the material. These delta electrons dramatically impact the response of detection systems when subject to ions with velocities typical of beams in modern fragmentation facilities. We test the accuracy of our model with data from Lutetium Yttrium Orthosilicate (LYSO:Ce), a common inorganic scintillator. We compare calculated light production and quenching factors with experimental data for heavy ions of varying mass and energy as well as make a quantitative estimate of the effects of $\delta$ rays on overall light output. The model presented herein will serve as a basic framework for further studies of scintillator response to heavy ions. Our results are crucial in planning future experiments where relativistic exotic nuclei are interacting with scintillator detectors.
\end{abstract} 

\section{Introduction}

Fragmentation facilities, such as the Radioactive Ion Beam Factory (RIBF) at RIKEN and the Facility for Rare Isotope Beams (FRIB), continue to increase the range of rare isotopes available for experimental studies. The need to expand the scope of radioactive decay studies, which requires a nano-second or sub-nanosecond response time scale, necessitated the use of scintillator arrays capable of ion-decay correlations \cite{XiaoPhysRevC100034315,CrawfordPhysRevLett129212501}, similar to those provided by conventional double-sided silicon strip detectors (DSSD), which until recently dominated the experiments \cite{was3abi,AIDA}. Inorganic-scintillator-based detector arrays were chosen for such applications due to their excellent timing,  large stopping power, and position sensitivity when coupled to position-sensitive light sensors \cite{YOKOYAMA201993,CrawfordPhysRevLett129212501,CRIDER_Thick_Scint,SIEGL,SINGH2025170239,OGUNBEKU,CHESTER,YOKOYAMA_One_Neutron,GRAY,LUBNA,COX,HENDERSON,NEUPANE,PELTIER2025139576}.
Detectors used at fragmentation facilities must be able to measure both low-energy decay and high-energy implantation events when the ion and decay occur in the same location. This energy deposition range spans from less than a hundred keV for decays to several GeV of total kinetic energy (TKE) for ion implantations. Without proper detector design, decay events may be too small to be detected, and implantation events may saturate the detector, resulting in lost information. For DSSD, the typical solution is to implement a dual-range electronic chain and let the high-gain (decay branch) saturate for implantation events. This method is not that straightforward for scintillators because of the limited dynamic range of the photomultipliers. Thus, it is crucial to accurately quantify the amount of light that will be generated in both decay and implantation events at the experiment design stage.

Practical implementation of these scintillation-based implant detectors is made feasible by an effect known as light quenching, whereby implanted ions produce less light than gammas or electrons at the same energy. Light quenching is due to increased ionization density in the ion's track, resulting in saturation of the scintillator's light output. The so-called quenching factor, $qf$, quantifies this effect through the ratio of actual ion TKE, $E$, to the light measured by the calibrated detector, $L$:

\begin{equation} \label{eq:qf}
    qf = \frac{E}{L}
\end{equation}

When assuming a linear response to $\gamma$-rays and electrons, scintillators can be calibrated using standard radioactive sources. In using such a calibration, we take the perfectly linear, unquenched response $L=E$ as a reference for comparing the quenching experienced by different ions. Light output measurements calibrated in this manner are sometimes denoted with units of MeVee or ``MeV electron equivalent." 

Estimating the light quenching from a scintillator detector is a crucial part of the experimental design process. While several models of the scintillation mechanism exist \cite{JBBirks_1951,MurrayMeyerInorganic,VOLTZ}, they are formulated in terms of differential quantities that are experimentally cumbersome to extract, making it difficult to determine responses for new scintillators across an extensive energy range or for many different nuclei. A model capable of predicting the light output as a function of the ion's TKE, known ion properties such as mass, and theoretically calculable quantities is needed. Another issue is that the evaluation of almost all of the existing models has been limited to somewhat low-$Z$ ions and/or low energy per nucleon $E/A$ \cite{MurrayMeyerInorganic,MurrayMeyerDelta,VOLTZ,HORN1992273,BECCHETTI197693,MCMAHAN,Koba2011ScintillationEO}, while it is the goal of radioactive ion beam facilities to push to high-$Z$ nuclei with high energies per nucleon. Therefore, we propose an approximate phenomenological approach to modeling light output and estimating quenching factors of heavy nuclei in Lutetium Yttrium Orthosilicate (LYSO:Ce, Lu$_{1.9}$Y$_{0.1}$SiO$_5$), a commonly-used inorganic scintillator. This method, based on the models of Birks \cite{JBBirks_1951,Koba2011ScintillationEO} and Meyer and Murray \cite{MurrayMeyerDelta}, is intended to serve as a basic framework for further studies of scintillator light production in the future.

\section{Theoretical Background}

The most well-known model of scintillator response to heavy ion beams was introduced by Birks \cite{JBBirks_1951}, who suggested that light quenching was the result of increased ionization density in the track of heavier incident ions. This, in turn, leads to a partial saturation of available luminescence centers inside the scintillator, decreasing the total light output. Birks' model expresses light output as a function of energy loss or stopping power:

\begin{equation}\label{eq:classic_birks_dLdx}
    \frac{dL}{dx} = \frac{a\frac{dE}{dx}}{1+b\frac{dE}{dx}}
\end{equation}

Here, $\frac{dL}{dx}$ is specific fluorescence, and $\frac{dE}{dx}$ is stopping power. The coefficients $a$ and $b$ are properties of the material and can be determined empirically by fitting $\frac{dL}{dx}$ as a function of $\frac{dE}{dx}$. Algebraic manipulation gives another version of the formula which is more convenient for our purposes \cite{Koba2011ScintillationEO}:

\begin{equation}\label{eq:classic_birks}
    \frac{dL}{dE} = \frac{a}{1+b\frac{dE}{dx}}
\end{equation}

$\frac{dL}{dE}$, referred to as scintillation efficiency, is a measure of how effectively the ion's energy is converted to light; in principle, it should take values between 0 and 1, provided that the detector has been calibrated. According to this elegantly simple model, the coefficients $a$ and $b$ should be unique for each scintillator and should approximately characterize its response to any ion at any energy. Though originally created for organic scintillators, the Birks formula seems to be a good approximation for the quenching effect in inorganic scintillators. One demonstration of this is the work of Koba et al. \cite{Koba2011ScintillationEO}, who performed scintillation efficiency measurements for protons,  $\alpha$ particles, $^{12}$C, and $^{40}$Ar in several inorganic scintillators, including LYSO. After using the Bethe-Bloch relation \cite{SalvatBetheBloch} to calculate stopping powers, they extracted the coefficients to a slightly modified version of the Birks formula for several inorganic scintillators. They then plotted the theoretically-predicted light output against experimental results, obtaining relatively good agreement. While more intricate models, such as that of Murray and Meyer \cite{MurrayMeyerInorganic}, have been proposed to account for low-$\frac{dE}{dx}$ effects in inorganic scintillators, the basic mechanism described by the Birks formula seems to explain relative light quenching for heavy ions with reasonable accuracy, as seen from both Koba et al.'s and others' results \cite{HORN1992273}. Despite this apparent success, Birks' model's assumption that ionization quenching is the only process contributing to nonlinear light production begins to break down for heavier ions $\left(Z \gtrsim 20\right)$ \cite{MurrayMeyerDelta,VOLTZ,MCMAHAN}. In this regime, we must address the issue of energetic secondary electrons.

\begin{figure}[]
    \centering
    \includegraphics[width=0.5\linewidth]{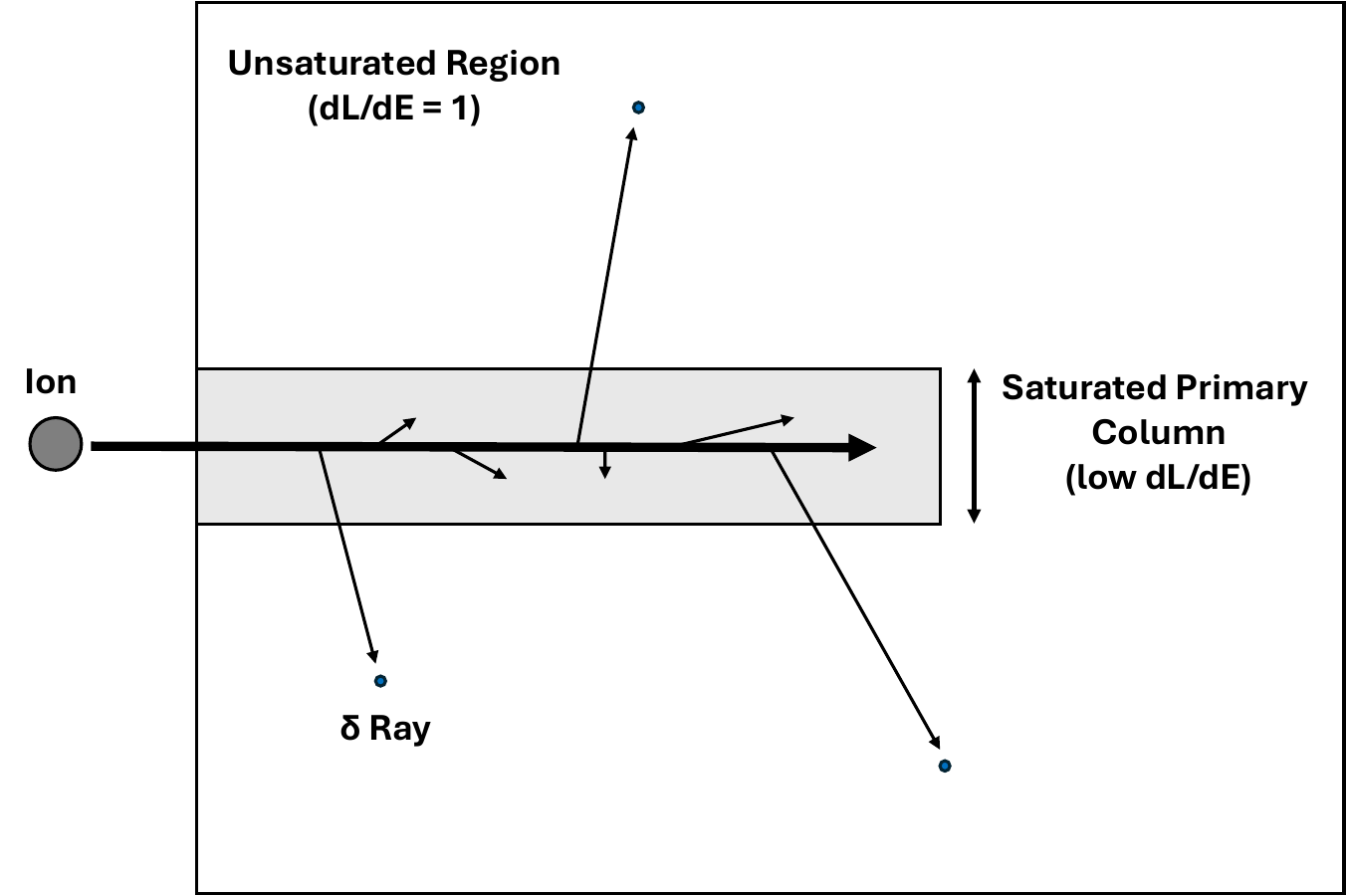}
    \caption{Qualitative picture of Meyer and Murray's model of the scintillation mechanism. As the ion travels through the scintillator, it rapidly loses energy within a narrow ``primary column'' of the material. High ionization density leads to a saturation of the available activator centers and, hence, lower light output in the column. A fraction of the lost energy contributes to the production of high-energy secondary electrons or $\delta$ rays. These $\delta$ rays are energetic enough to leave the primary column and enter the unsaturated parts of the crystal, scintillating with the characteristic high efficiency of electrons.}
    \label{fig:scint_mech}
\end{figure}

In a follow-up to their original paper on the inorganic scintillation mechanism, Meyer and Murray \cite{MurrayMeyerDelta} examined how the production of high-energy secondary electrons, known as $\delta$ rays, could contribute to scintillator light output. While they focused on activated alkali halides such as NaI(Tl), their analysis can be extended to other activated inorganic scintillators and likely even organic scintillators. A qualitative depiction of their model of the scintillation mechanism is shown in Fig. \ref{fig:scint_mech}. As a heavy ion traverses the scintillator, it loses energy rapidly through a series of collisions, resulting in a narrow column of very high ionization density. The relatively few light-producing activator sites within this primary column begin to saturate due to the large amount of ionization. This saturation leads to quenching, as in Birks' model. As expected, increasing the incident ion's mass further increases the ionization density and results in lower light output in the column. However, unlike in the Birks model, some fraction of the ion's energy will be imparted to $\delta$ rays. These $\delta$ rays have sufficient energy to escape the saturated primary column and produce light in an untouched region of the material with a scintillation efficiency close to 1, much like regular gamma and beta particles. Meyer and Murray argued that $\delta$ rays were the reason for small differences in scintillation efficiency (which they dubbed ``fine structure") between ions of similar mass and different $Z$. Since their paper, there have been relatively few attempts to include $\delta$ ray effects in models of scintillator response \cite{BECCHETTI197693,MCMAHAN,VOLTZ}. It might seem, at least for the ions that have been examined in these and other studies, that the ionization quenching mechanism is sufficient to characterize the relative light output of different ion species, as seen in the results of Koba et al. \cite{Koba2011ScintillationEO}. Presumably, the fraction of absolute light contributed by $\delta$ rays was small enough to not significantly impact the relative light output predicted by the Birks formula.

As ion mass is increased further, however, we can no longer ignore $\delta$ rays. The very severe ionization quenching for such ions means that the $\delta$ ray mechanism could dominate the light output. This is apparent from the light output measurements reported here. The data were collected during a recently performed experiment at RIBF RIKEN. Neutron-deficient isotopes around $^{100}$Sn were produced by 345 MeV/u $^{124}$Xe primary beam and were implanted into a 0.6 mm unsegmented LYSO crystal. A segmented light guide was coupled between the LYSO crystal and a Hamamatsu H12700B-10 multi-anode photomultiplier tube, with position sensitivity.  This detector was based on previous inorganic scintillator-based implant detectors \cite{YOKOYAMA201993,XiaoPhysRevC100034315,SINGH2025170239}.  Ions measured in this experiment were implanted with energies in the range of about 1-10 GeV and atomic number $Z \geq$ 50. The very high-$Z$ exotic ions, such as $^{108}$Te ($Z$ = 52), exhibited light output comparable to that measured by Koba et al. \cite{Koba2011ScintillationEO} for $^{40}$Ar at the same energy, as shown in the top panel of Fig. \ref{fig:light_and_qf}; according to a classic ionization quenching model such as Birks' (represented by the dotted brown line in Fig. \ref{fig:light_and_qf}), the light output for $^{108}$Te should have been an order of magnitude smaller due to its significantly higher stopping power in this energy range. This would suggest that the total light output is due almost exclusively to the $\delta$ ray contribution. The simplistic ionization quenching model is clearly not adequate in this regime. Luckily, as will be shown, the Birks formula may still be salvageable, provided some correction is made for $\delta$ ray effects. The development and testing of a new model are outlined in the following sections.

\section{Developing a Model With $\delta$ Rays}

Meyer and Murray's formulation treats the light output contributions from the saturated primary column and $\delta$ rays separately:

\begin{equation}\label{eq:murray_meyer_1}
    \left(\frac{dL}{dE}\right)_{t} = (1-F)\left(\frac{dL}{dE}\right)_{p} + F\left(\frac{dL}{dE}\right)_{\delta}
\end{equation}

Where

\begin{equation}
    F = \frac{\left(\frac{dE}{dx}\right)_{\delta}}{\left(\frac{dE}{dx}\right)_{t}}
\end{equation}

Here, $\left(\frac{dL}{dE}\right)_{\delta}$ and $\left(\frac{dL}{dE}\right)_{p}$ refer to the scintillation efficiencies of the $\delta$-ray and primary column light production mechanisms, respectively. $\left(\frac{dL}{dE}\right)_{t}$ is the total scintillation efficiency. Similarly, $\left(\frac{dE}{dx}\right)_{\delta}$ refers to the energy loss to $\delta$ rays along the ion track, while $\left(\frac{dE}{dx}\right)_{t}$ refers to the total energy loss. Hence, $F$ is just the fraction of the ion's energy loss imparted to $\delta$ rays. We can further simplify by assuming that the secondary electrons scintillate with an efficiency of about 1 \cite{MurrayMeyerDelta}:

\begin{equation}\label{eq:murray_meyer_2}
    \left(\frac{dL}{dE}\right)_{t} \approx (1-F)\left(\frac{dL}{dE}\right)_{p} + F
\end{equation}

As noted by Meyer and Murray, we do not have a simple, closed-form expression for $F$, so we must resort to approximations. Before looking at these, however, we need to define a few terms. Assume that the ion enters the scintillator perpendicular to its surface and travels in a straight line; this line corresponds to the central axis of the primary column. $\epsilon_0$ is the total kinetic energy of a $\delta$ ray, $r$ is its radial distance from the primary column axis, and $\theta$ is the angle at which it is ejected with respect to the primary column axis. Meyer and Murray observed that the $\delta$ ray range $R_p$ can be approximated by a power law relation:

\begin{equation}
    R_p = R_{0}\epsilon_{0}^{n}
\end{equation}

We can obtain values for $R_{0}$ and $n$ by fitting electron range curves from ESTAR \cite{ESTAR}. For LYSO, it was found that $R_{0}$ is 1.37 mg/(cm$^2$keV$^{n}$) and $n$ is 0.915 if $R_p$ is in units of mg/cm$^2$ and $\epsilon_0$ is in units of keV. We will continue to use units of mg/cm$^2$ for distances and keV for energies throughout our analysis. We define the primary column radius $r_{c}$ as the minimum distance that a $\delta$ ray needs to travel to leave the saturated primary column and scintillate with high efficiency. As Meyer and Murray point out, $r_c$ is somewhat of an abstraction; physically, the primary column would not have a clearly defined radius. It would also exhibit a small dependence on $\frac{dE}{dx}$. It is simplest, however, following in their footsteps, to treat $r_c$ as a constant fitting parameter \cite{MurrayMeyerDelta}. The last quantity we need to define is the effective charge $z^{*}$ of the ion, given approximately by \cite{ZieglerStopping,FUNKHOUSER2002377,AnthonyLanfordZeff}

\begin{equation}
    z^{*} = z\left[1-exp\left(\frac{-v}{z^{2/3}\alpha c}\right)\right]
\end{equation}

where $z$ is the atomic number, $v$ is the ion velocity, $\alpha$ is the fine structure constant, and $c$ is the speed of light. Meyer and Murray derived three different estimates of $F$ for the inorganic scintillator NaI(Tl) using the Rutherford distribution. Their approach was to find an approximate expression for the energy loss to $\delta$ rays $\left(\frac{dE}{dx}\right)_{\delta}$ and then normalize by the total stopping power $\left(\frac{dE}{dx}\right)_{t}$. One of these estimates is based on the assumption that $\delta$ rays are emitted from the primary column in an isotropic fashion:

\begin{multline}\label{eq:F_formula}
    \left(\frac{dE}{dx}\right)_{\delta} =\frac{3.08\times10^5}{(E/A)}z^{*2}\int_{\epsilon_{0}^{min}}^{\epsilon_{0}^{max}}\frac{d\epsilon}{R_{p}^{3}\epsilon} \int_{r_c}^{R_p}\left(1-\frac{r}{R_p}\right)r^{2}dr \int_{arcsin(r_{c}/r)}^{\pi/2}\left(1-\frac{r_c}{r sin\theta}\right)^{1/n} \mathrm{sin}\theta d\theta
\end{multline}

The upper and lower limits of the energy integral, $\epsilon_{0}^{min}$ and $\epsilon_{0}^{max}$ are given by $(r_c/a)^{1/n}$ and $4m(E/M) \approx (4/1822)(E/A)$, respectively. The former is the minimum $\delta$ ray energy needed to travel outside the primary column, and the latter is the maximum $\delta$ ray energy that can be obtained from the Rutherford scattering formula \cite{MurrayMeyerDelta}. $M$ is the mass of the ion and $m$ is the electron mass. For more details on the derivation of Eq. (\ref{eq:F_formula}), the reader is referred to Meyer and Murray's work. To obtain a complete expression for the total scintillation efficiency, we can approximate the primary column scintillation efficiency $\left(\frac{dL}{dE}\right)_{p}$ with the Birks formula, Eq. (\ref{eq:classic_birks}), as some authors have done in previous works \cite{MCMAHAN,VOLTZ}. Hence, our model becomes

\begin{equation}
    \left(\frac{dL}{dE}\right)_t = (1-F')\frac{a}{1+b(1-F')\left(\frac{dE}{dx}\right)_t}+F'
\end{equation}

Where

\begin{equation}
    F' = cF
\end{equation}

With $F$ given by Eq. (\ref{eq:F_formula}). We include the additional scaling factor $c$ in front of $F$ as a fitting parameter to account for differences in material. There is another clever trick we can use to simplify our analysis. As noted previously, we desire a model which directly gives us total light output $L$ rather than scintillation efficiency $\left(\frac{dL}{dE}\right)_t$, since $L$ can be easily measured during experiments without modifying the detector. It turns out that we can achieve this by simply fitting the integral of the scintillation efficiency over the energy $E$ of the ion:

\begin{equation}\label{eq:complete_model}
    L = \int_{0}^{E}\left[(1-F')\frac{a}{1+b(1-F')\left(\frac{dE'}{dx}\right)_t}+F'\right]dE'
\end{equation}

It is important to point out again that $L$ is the normalized total light output i.e. the ion energy as measured by the detector; in this work, a simple $^{137}$Cs calibration (in which the light output is only scaled) is used so that the nonlinearity of the response is preserved. The other quantities relevant to Eq. (\ref{eq:complete_model}) may be obtained easily through measurement or theoretical calculation. Ion total kinetic energy $E$ may be either measured or estimated using the LISE++ code \cite{TARASOV20084657}. The stopping power $\frac{dE}{dx}$ corresponding to energy $E$ may be calculated analytically using the Bethe-Bloch formula with corrections \cite{SalvatBetheBloch} or a software package such as the Stopping and Range of Ions in Matter (SRIM) \cite{ZIEGLER20101818}; we have used the latter in this work. Thus, the only unknown quantities are $a$, $b$, $c$, and $r_c$. These will be our fitting parameters.

\section{Testing the Model with LYSO}

\begin{figure}[]
    \centering
    \includegraphics[width=0.5\linewidth]{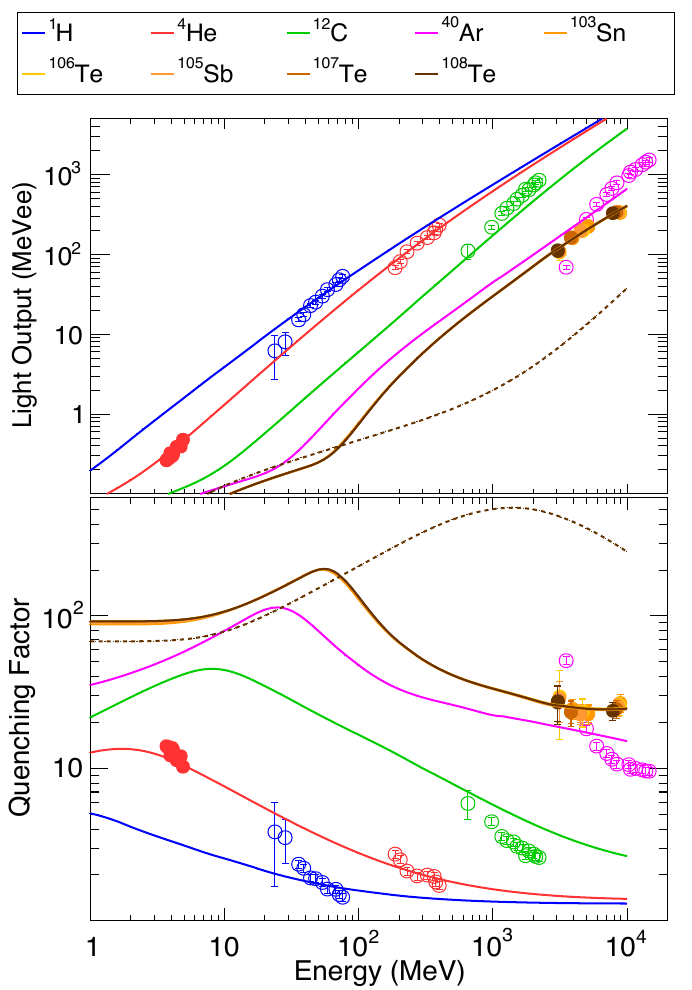}
    \caption{Light output (top) and quenching factor (bottom) as a function of ion energy. The solid light output curves were calculated with Eq. (\ref{eq:complete_model}), and the quenching factor curves were calculated with Eq. (\ref{eq:qf}). Data from Koba et al. \cite{Koba2011ScintillationEO} is depicted with open markers, and data from the RIKEN experiment is depicted with solid markers. Light output is expressed in units of MeVee or ``MeV electron equivalent." This means the energy measured by the detector after calibration with $\gamma$-ray sources. The dotted line represents the results from the Birks formula Eq. (\ref{eq:classic_birks}) for $^{108}$Te. As discussed, Birks formula significantly underestimates the light output of these very heavy ions.}
    \label{fig:light_and_qf}
\end{figure}

The two previously-mentioned sets of LYSO light production data (shown in the top panel of Fig. \ref{fig:light_and_qf}) were used to obtain the coefficients to Eq. (\ref{eq:complete_model}). The first set of data was taken from Koba et al. \cite{Koba2011ScintillationEO} and extracted using a plot digitization software. We have included their light production data for protons, $\alpha$ particles, $^{12}$C, and $^{40}$Ar. It is worth noting that these measurements were taken at relatively large $E/A$, between 20 and 500 MeV per nucleon. Since it was possible to digitize the error bars of only a few light production measurements, the others were assumed to have an error of 6\%. Conveniently, Koba et al. normalized their light production measurements with a $^{137}$Cs source. The second set of light production data, taken from the recent RIKEN experiment, includes several high-$Z$ ions: $^{103}$Sn, $^{106}$Te, $^{105}$Sb, $^{107}$Te, and $^{108}$Te. The TKE for each of the ions was obtained from calculations with the LISE++ code \cite{TARASOV20084657} using known energy based on the magnetic rigidity of the fragment separator. Once again, these measurements were performed at high $E/A$ ranging from 20 to 100 MeV per nucleon. A few lower-energy $\alpha$ particles of known energy from decays were also included so that the data would have a wider range of $E/A$. These light production measurements were similarly normalized with a $^{137}$Cs source. Stopping powers were obtained as a function of energy by interpolating SRIM \cite{ZIEGLER20101818} tables. The ROOT package was used to fit Eq. (\ref{eq:complete_model}) to the experimental data. To simplify and speed up the fitting code, we reduced the fit from three variables ($E$, $Z$, $A$) to two ($E$ and $Z$) by noting that $A \approx 2Z$ for this set of ions. This is not the case in general but suffices for our present analysis. The fit coefficients are listed in Tab. \ref{tab:coefficients}. Note that $r_c$ is given in units of $\mu$m. Interestingly, the $r_c$ of 1.4 $\mu$m extracted from fitting is about 30-40 times larger than the 400 \r{A} used by Meyer and Murray \cite{MurrayMeyerDelta}. This makes sense because the ions examined here are much heavier than those investigated in their work. Hence, the ion's energy loss is greater. While the model assumes a constant radius, such a clear fixed boundary will not exist in reality. We might stipulate that, as energy loss increases, the energy will be spread over an increasingly large radius. Our $r_c$ just represents an approximation for ions of varying mass. Though Meyer and Murray suggested that $r_c$'s dependence on $\frac{dE}{dx}$ is very weak, this might not be true once the model is extended to very large masses.



\begin{table}[]
    \centering
    \caption{Coefficients to Eq. (\ref{eq:complete_model}), obtained from fitting.}
    \begin{tabular}{c c c c}
        \hline
        \hline
        $a$ & $b$ (mg/keV$\cdot$cm$^2$) & $c$ & $r_c$ ($\mu$m) \\
        \hline
        0.795 & 0.0153 & 0.190 & 1.41  \\
        \hline
        \hline

    \end{tabular}
    \label{tab:coefficients}
\end{table}

The calculated light output and quenching factor are plotted as a function of energy with experimental values in the top and bottom panels of Fig. \ref{fig:light_and_qf}, respectively. Eq. (\ref{eq:qf}) was used to obtain quenching factors from both theoretical and experimental data. The calculated and experimental light output and quenching factors exhibit excellent agreement for very heavy ions and $\alpha$ particles. The inclusion of $\delta$ ray effects in the model leads to much greater accuracy for very heavy ions than could be achieved by an ionization quenching model alone. For comparison, the predictions of the Birks formula Eq. (\ref{eq:classic_birks}) are shown for $^{108}$Te as a dotted line; while the Birks formula underestimates the light output of $^{108}$Te and other very heavy ions by an order of magnitude, the improved model matches the measurements quite nicely. The good results for $\alpha$ particles over a wide energy range confirm that $\delta$ ray effects are largely a function of $E/A$, as predicted by Meyer and Murray. The agreement is not as good for protons, $^{12}$C, or $^{40}$Ar, though the calculated values are all within an order of magnitude of the experimental values. While it might be possible to obtain better agreement with a more sophisticated model, the simple model used here correctly predicts most of the qualitative features of the response and should be an acceptable starting point for future analysis.

\begin{figure}[]
    \centering
    \includegraphics[width=0.5\linewidth]{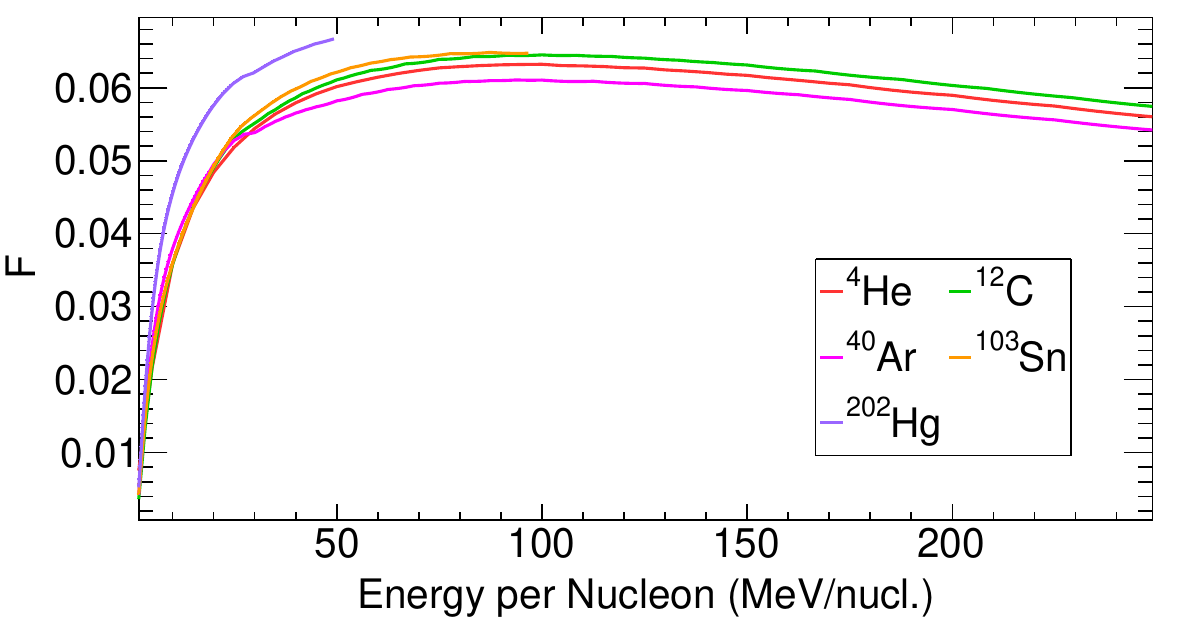}
    \caption{Fraction of ion energy energy loss $F$ imparted to $\delta$ rays as a function of energy per nucleon. These curves were calculated for energies up to 10 GeV using Meyer and Murray's isotropic emission approximation, Eq. (\ref{eq:F_formula}). We have restricted the energy here because some of the assumptions underlying Eq. (9) (i.e. the Rutherford scattering model, which is non-relativistic) begin to break down at very high energies.}
    \label{fig:delta_ray_fraction}
\end{figure}

\begin{figure}[h]
    \centering
    \includegraphics[width=0.5\linewidth]{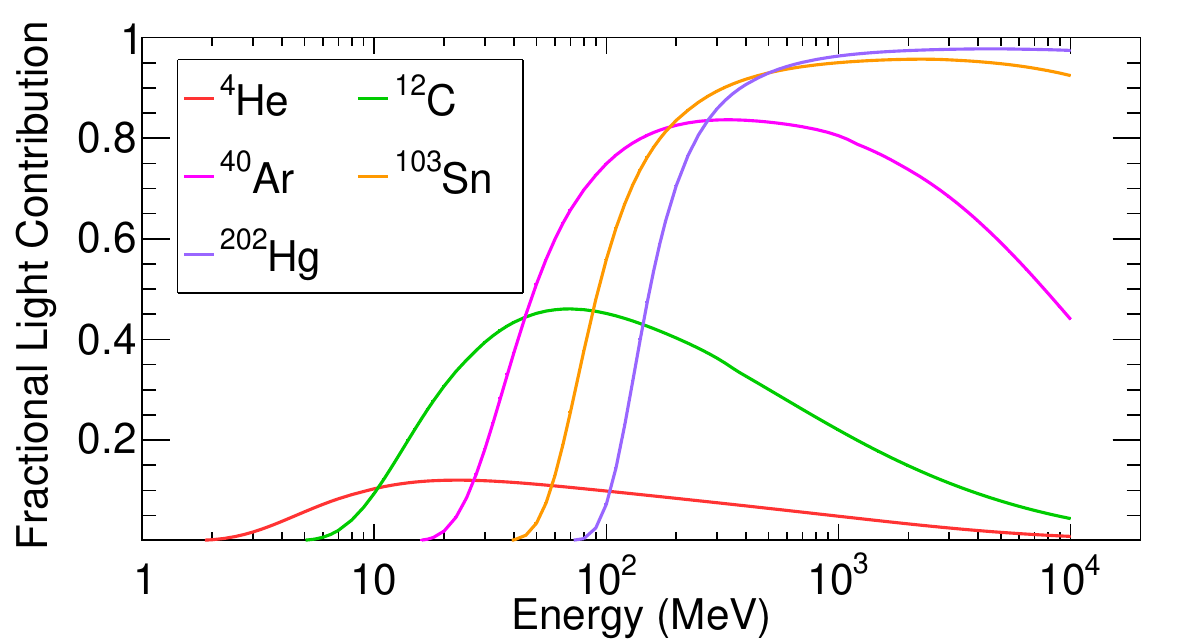}
    \caption{Calculated fraction of light output due to $\delta$ rays. Note that light output is almost entirely determined by $\delta$ rays for very heavy ions.}
    \label{fig:light_fraction}
\end{figure}

Using Eq. (\ref{eq:F_formula}), we can calculate the $\delta$ ray fraction $F$ as a function of energy per nucleon, as shown in Fig. \ref{fig:delta_ray_fraction} for a handful of ions with energies up to 10 GeV. $^{202}$Hg ($Z=80$) is included to show what the model predicts for much larger $Z$. When plotted as a function of energy per nucleon, $F$ is roughly the same for $^{4}$He and heavier ions, although it is somewhat larger for $^{202}$Hg. It is easy to find the fraction of light due to $\delta$ rays by dividing the contribution from the second term in Eq. (\ref{eq:complete_model}) by the total light output. The result is shown as a function of energy in Fig. \ref{fig:light_fraction}. These two plots explain why the $\delta$ ray effect is so much more pronounced for very heavy ions. Ionization quenching continues to worsen in the primary column as ion mass increases, while the scintillation efficiency of $\delta$ rays remains close to 1. The net result is that the light output of ions heavier than $^{40}$Ar is due almost entirely to $\delta$ rays. This appears to place a lower limit on the quenching mechanism and, by extension, the expected light output of very heavy ions.

\section{Conclusions}

We have shown a simple empirical framework for estimating the light output of inorganic scintillators based on the models of Birks, Meyer, and Murray. Unlike these models, it is formulated in terms of total light output $L$ instead of scintillation efficiency $\frac{dL}{dE}$. This eliminates the need for direct scintillation efficiency measurements. The agreement of the model with experimental data for LYSO was relatively good over a large range of masses and energies; differences, even in the worst case, did not exceed an order of magnitude. We calculated the $\delta$ ray fraction curves for various ions and found them to be somewhat similar as a function of $E/A$. Additionally, we estimated the relative contribution of $\delta$ rays to total light output. For ions heavier than $^{40}$Ar, light output appears to be dominated by $\delta$ rays, indicating a limit on the light quenching effect. 

There is good reason to believe that geometry-dependent effects and light-readout-related considerations also become important when operating over a wide range of masses and energies, but analysis of these factors is beyond the scope of this work. A rigorous treatment will be deferred until more data is available. Future studies should include a more extensive range of ion energies and masses. They should also, as far as possible, employ the same detector geometry across measurements and include corrections for relevant external effects. The simplified model and discussion presented herein may be the first step in developing better implant detectors and adequately characterizing their response.


\section*{Acknowledgments}

We would like to express our gratitude to the entire RIBF operations team for ensuring reliable beam delivery throughout the experiment. This work is partly funded by the U.S. Department of Energy, Office of Science, Office of Nuclear Physics under Contracts No. DE-AC05-00OR22725 (ORNL) and No. DE-FG02-96ER40983 (UTK) as well as the Polish National Science Center under Grant No. 2020/39/B/ST2/02346 (UW). Additionally, this work was sponsored by the Stewardship Science Academic Alliances program through DOE Award No. DE-NA0003899 (UTK) and NSF Major Research Instrumentation Program Award No. 1919735 (UTK).


\bibliographystyle{unsrt}
\bibliography{references}

\end{document}